\documentclass[a4paper, 11pt]{article}
\pdfoutput=1
\usepackage{jcappub}
\usepackage[utf8]{inputenc}

\usepackage{epsfig}
\usepackage{url}

\usepackage[normalem]{ulem} 

\usepackage{latexsym}
\usepackage{epsfig}
\usepackage{amsmath}
\usepackage{amssymb}
\usepackage{wasysym}
\usepackage{graphicx}
\usepackage{verbatim}
\usepackage{enumerate,mdwlist}
\usepackage[titletoc]{appendix}
\usepackage{amsfonts}
\usepackage{tikz} 
\usetikzlibrary{calc}
\usepackage[export]{adjustbox}
\usepackage[caption=false]{subfig}

\usepackage{units}

\usepackage[normalem]{ulem}

\newcommand{\erf}{\mbox{erf}}

\newcommand{\tr}{\mbox{tr}}

\newcommand{\ba}{\begin{eqnarray}}
\newcommand{\ea}{\end{eqnarray}}
\newcommand{\be}{\begin{equation}}
\newcommand{\ee}{\end{equation}}
\newcommand{\bfa}{{\bf a}}
\newcommand{\bfn}{{\bf n}}
\newcommand{\bfs}{{\bf s}}

\newcommand{\al}{\alpha}

\definecolor{grey}{rgb}{0.4,0.4,0.4}
\definecolor{dullmagenta}{rgb}{0.4,0,0.4}
\definecolor{darkblue}{rgb}{0,0,0.4}
\definecolor{midblue}{rgb}{0,0,0.5}
\definecolor{midred}{rgb}{0.5,0,0}
\definecolor{orange}{rgb}{1,0.5,0}
\definecolor{lightbrown}{rgb}{0.75,0.5,0.25}
\definecolor{tan}{cmyk}{0.14,0.42,0.56,0}
\definecolor{djunglegreen}{cmyk}{0.99,0,0.52,0}
\definecolor{lightgreen}{rgb}{0,1,0}
\definecolor{olivegreen}{cmyk}{0.64,0,0.95,0.40}
\definecolor{midgreen}{rgb}{0.0,0.675,0.0}
\definecolor{darkgreen}{rgb}{0,0.5,0}

\usepackage[all]{xy} 
\usepackage{amsfonts}

\DeclareMathOperator{\arctanh}{arctanh}

\title{$\alpha$-attractor dark energy in view of next-generation cosmological surveys}

\author[a,b,c,d]{Carlos Garc\'ia-Garc\'ia}
  \emailAdd{carlosgarcia@iff.csic.es}
\author[b,a]{Pilar Ru\'iz-Lapuente}
  \emailAdd{pilar@icc.ub.edu}
\author[d]{David Alonso}
  \emailAdd{david.alonso@physics.ox.ac.uk}
\author[c,e]{M.~Zumalac\'arregui}
  \emailAdd{miguelzuma@berkeley.edu}

\affiliation[a]{Instituto de Física Fundamental, Consejo Superior de Investigaciones
Científicas, c/. Serrano 121, E–28006, Madrid, Spain}
\affiliation[b]{ Institut de Ci\`{e}ncies del Cosmos (UB–IEEC), c/. Martí i
 Franqués 1, E–08028, Barcelona, Spain}
\affiliation[c]{Berkeley Center for Cosmological Physics and University of California at Berkeley, CA94720, USA}
\affiliation[d]{Department of Physics, University of Oxford, Denys Wilkinson
  Building, Keble Road, Oxford OX1 3RH, United Kingdom}
\affiliation[e]{Institut de Physique Th\' eorique, Universit\'e  Paris Saclay 
CEA, CNRS, 91191 Gif-sur-Yvette, France}

\abstract{%
  The $\alpha$-attractor inflationary models are nowadays favored by CMB Planck
  observations. Their similarity with canonical quintessence models motivates
  the exploration of a common framework that explains both inflation and dark
  energy. We study the expected constraints that next-generation cosmological
  experiments will be able to impose for the dark energy $\alpha$-attractor
  model. We systematically account for the constraining power of SNIa from
  WFIRST, BAO from DESI and WFIRST, galaxy clustering and shear from LSST and
  Stage-4 CMB experiments.  We assume a tensor-to-scalar ratio, $10^{-3} < r <
  10^{-2}$, which permits to explore the wide regime sufficiently close, but
  distinct, to a cosmological constant, without need of fine tunning the
  initial value of the field. We find that the combination S4CMB + LSST + SNIa
  will achieve the best results, improving the FoM by almost an order of
  magnitude; respect to the S4CMB + BAO + SNIa case. We find this is also true
  for the FoM of the $w_0 - w_a$ parameters. Therefore, future surveys will be
  uniquely able to probe models connecting early and late cosmic acceleration.
}

\begin{document} 
\maketitle
 
\section{Introduction}

The standard model of cosmology relies on two epochs of accelerated expansion.
A first inflationary phase in the very early universe leading to a very
homogeneous, isotropic and spatially flat with a near scale invariant spectrum
of curvature perturbations \cite{Linde:2014nna,Akrami:2018odb}.  The second
acceleration era, when dark energy (DE) dominates the energy density in the
late universe, is necessary to explain observations of type Ia supernovae
(e.g. Refs.~\cite{Riess:1998cb,Perlmutter:1998np,Abbott:2018wog}), the cosmic
microwave background (CMB) (e.g. Refs.~\cite{Spergel:2013tha,Aghanim:2018eyx})
and the large-scale structure (LSS) in the matter distribution
(e.g. Ref.~\cite{Alam:2016hwk}). An ambitious observational program aims at elucidating
the physics behind inflation and DE. 

In this context, the dark-energy $\alpha$-attractor model
\cite{Linder:2015qxa} is one of the models that try to describe both
accelerated expansions in a common framework.  These models typically have a
scalar field in a potential with two plateaus that allow for a slow roll at
early times, which produces inflation, and a freezing behavior at late times,
that yields a cosmological constant-like expansion
\cite{Dimopoulos:2017zvq,Dimopoulos:2017tud,Akrami:2017cir,Casas:2017wjh}.
In addition, there are models that would produce dark energy from a symmetry
breaking mechanism
\cite{Dimopoulos:2018eam,Casas:2018fum,Camargo-Molina:2019faa}.
However, there are other studies that try to study the connection of the late and early
Universe, but focus only on the late time cosmology. Among them there are
those describing dark energy as quintessence, which base their Lagrangian on
an $\alpha$-attractor model
\cite{Linder:2015qxa,Shahalam:2016juu,Bag:2017vjp,Garcia-Garcia:2018hlc,Cedeno:2019cgr}, or
those which study the relation between them and $f(R)$ gravity, from
extensions of the Starobinsky $R^2$ gravity \cite{Starobinsky:1980te}, as in
Refs.  \cite{Odintsov:2016vzz,Miranda:2017juz}. Others, instead, use the
$\alpha$-attractors as source of dark matter \cite{Mishra:2017ehw}.

During inflation, the $\alpha$-attractors class of models shines as a group of
models able to reproduce the observations, which strongly support concave
potential models. CMB Planck sets tight constraints on the
tensor-to-scalar ratio, $r$, with $r < 0.11$ (at $95\%$ CL) and the spectral
index, $n_s$, with $n_s=0.9649\pm 0.0042$ (at $95\%$ CL), favoring
slow-roll models with a concave potential ($V(\phi)'' < 0$)
\cite{Akrami:2018odb}, as was already anticipated by WMAP results
\cite{Hinshaw:2012aka}. In this context, the $\alpha$-attractor models are
able to give the correct predictions thanks to the fact that, for $N$ e-folds
\cite{Galante:2014ifa},
\begin{equation}
  n_s =  1 - 2 N^{-1} \quad \mbox{ and } \quad r = 12 \alpha N^{-2}\,,
  \label{eq:ns}
\end{equation}
where $\alpha$ is a parameter shared by all models in this class and is
present in their Lagrangian, whose canonical expression is given by
\begin{equation}
{\mathcal L}=\sqrt{-g}\,\left[\frac{1}{2}M_P^2 R- \frac{1}{2} (\partial\phi)^2
-\alpha f^2(x)\right]\ ,
\label{eq:L}
\end{equation}
where $x = \tanh(\phi/\sqrt{6\alpha})$. The fact that their Lagrangian is the
same as the one for canonical quintessence dark energy models is exploited to
connect both inflation and dark energy with the same scalar field.

The $\alpha$-attractor models are connected with fundamental theories
with various fields with local conformal (i.e. rescaling) invariance.
  This symmetry allows to rewrite the original Lagrangian
  as a single-field one~\cite{Galante:2014ifa,Kallosh:2013hoa},
\begin{equation}
  {\mathcal L}=\sqrt{-g}\,\left[\frac{1}{2}M_P^2 R-\frac{\alpha}{(1-\varphi^2/6)^2}\,\frac{1}{2} (\partial\varphi)^2
  -\alpha f^2\left(\frac{\varphi}{\sqrt{6}}\right)\right]\ .
  \label{eq:L-original}
\end{equation}
Here, $g$ is the metric and $R$ the Ricci scalar, $M_P$ is the Planck mass,
and $\alpha f^2$ is the potential function dependent on the field $\varphi$
which is measured in $M_P$ units. The second order pole in the kinetic term is
the reason behind the common predictions for $n_s$ and $r$ (Eq.~\ref{eq:ns}).
Finally, in order to obtain its canonical version, one needs to define
$\phi=\sqrt{6\alpha}\,\arctanh(\varphi/\sqrt{6})$. In this way, one also pushes
the boundaries of the connected region ($\varphi \in (-\sqrt{6}, \sqrt{6})$)
to infinity ($\phi \in (-\infty, \infty)$). 

In this paper, we will extend the previous work of
Ref.~\cite{Garcia-Garcia:2018hlc} that studied the phenomenology and the
observational constraints of a generalized $\alpha$-attractor dark energy
model, based on Starobinsky $R^2$ inflationary Lagrangian
\cite{Starobinsky:1980te}. It was shown that the generalized Starobinsky-model
had an infinite $\Lambda$CDM-like region which made that, imposing only late
time observational constraints (and a true model close to $\Lambda$CDM),  one
could only have lower bounds on the initial position of the field and $\alpha$
and the requirement on the exponents of being of the same order, so that the
field slow rolls. 

We will systematically study how future observations will affect
the constraints on the model's parameters. Next spectroscopic experiments
generation will reduce the relative error on the angular diameter distance
and the Hubble parameter to order of a few percents, while deepening up to
$z \sim 3$ \cite{Aghamousa:2016zmz,Font-Ribera:2013rwa}. Baryonic Acoustic
Oscillations (BAO) measurements will also significantly
improve their accuracy, what will be reflected on the parameter constraints
\cite{Garcia-Garcia:2018hlc}. On the other hand, Stage 4-CMB experiments are
expected to measure the tensor-to-scalar ratio to order $\sigma(r) \sim 0.001$
\cite{Abazajian:2013vfg}. Lowering the $r$ upper-bound below $ \sim 0.01$
might be able to constrain $\alpha$ through Eq. \ref{eq:ns} and, in turn, the
initial value of the field. Finally, from the Large Synoptic Survey Telescope
\cite{Abell:2009aa}, a photometric experiment, we will take into account their
predictions for galaxy clustering and shear measurements, which will effectively
constrain cosmological parameters by means of precise measurements of the matter
power spectrum at different redshifts.

It is important to note that some of these next-generation experiments will
overlap, allowing to beat cosmic variance when cross-correlations are taken
into account \cite{Seljak:2008xr}. In order to take advantage of this
piece of information, we will use the multi-tracer formalism
as described in Ref.~\cite{Alonso:2015sfa}, which is reviewed in Section \ref{S:Fisher}.

In Section \ref{S:model}, we will introduce the $\alpha$-attractor dark
energy model and summarize its properties. In Section \ref{S:probes}, we will
briefly summarize the models used to describe each observational probe that enter our forecast:
Stage-4 CMB experiments, the Large Synoptic Survey Telescope
\cite{Abell:2009aa}, DESI \cite{Aghamousa:2016zmz} and WFIRST
\cite{Spergel:2013tha}.  In Section \ref{S:Fisher}, we will review the
multi-tracer Fisher formalism and the computational tools that carry out the
computations. In Section \ref{S:results}, the forecasted constraints will be shown
and analyzed.  Finally, in section \ref{S:conclusions}, we will conclude.

\section{The generalized $\alpha$-attractor model}
\label{S:model}
The generalized $\alpha$-attractor model was first proposed in Ref.
\cite{Linder:2015qxa} and further studied in Refs.
\cite{Shahalam:2016juu,Garcia-Garcia:2018hlc}. It generalizes the
Starobinsky inflationary model \cite{Starobinsky:1980te}, freeing the
potential's exponents and amplitude:
\begin{equation}
  V(x)=\al c^2\,\frac{x^p}{(1+x)^{2n}} = \al c^2 2^{-2n}(1-y)^p (1+y)^{2n-p} \ , 
  \label{eq:V}
\end{equation}
where $c$, $p$, $n$ are constant parameters, $x = \tanh(\phi/\sqrt{6\alpha})$
and $y \equiv e^{-2\phi/\sqrt{6 \alpha}}$. The Starobinsky
model is obtained with $\alpha = 1$, $p = 2$, $n = 1$ \cite{Whitt:1984pd,
Maeda:1987xf, Barrow:1988xi}, in natural units, i.e.\ reduced Planck
mass $M_P = 1$ and speed of light, $c = 1$. We assume a flat Universe as a
consequence of inflation.

In this section, we will summarize the properties of this model, already
studied in Ref. \cite{Linder:2015qxa,Garcia-Garcia:2018hlc}. We will use the
scaled field variable, $\psi \equiv \phi/\sqrt{6}$, introduced in Ref.
\cite{Garcia-Garcia:2018hlc}, as it better reflects what is the determining
quantity in $V(x)$. Let us list them bellow:
\begin{itemize}
  \item The potential Eq. \ref{eq:V} has two limits -- it has a power
    law regime at low $\psi$ and a plateau at large $\psi$
    \cite{Linder:2015qxa}:
    \begin{eqnarray}
      V(|\psi|\ll \sqrt{6}) &\approx& \alpha c^2 6^{-p/2}\,\psi^p \ ,
      \label{eq:Vflow}\\
      V(\psi \gg \sqrt{6}) &\approx& \al c^2\,2^{-2n}\,
      \left[1-2(p-n)\,e^{-2\psi/\sqrt{6}}\right] \\
        \xrightarrow[\psi \to \infty]{} && \frac{\alpha c^2}{2^{2n}} \ .
      \label{eq:Vfgtr}
    \end{eqnarray}
  \item Viable models are of the thawing class and $\Delta \psi \equiv
      \psi_0 - \psi_i$ (the field excursion, i.e. the difference between its
      initial, $\psi_i$ and today's, $\psi_0$, values), can be approximated as
      $\Delta \psi \sim \sqrt{1+w_0}/\sqrt{\alpha}$. In addition, it can be
      shown that $1+w_0\sim 1/\alpha$, yielding $\Delta \psi \sim 1/\alpha$
    \cite{Garcia-Garcia:2018hlc}. 
  \item If $n<p$, the field always decreases \cite{Linder:2015qxa} and
    its velocity is inversely related to its initial value ($\psi_{ini}$).
    Also, $p$ determine the transition regime slope
    \cite{Garcia-Garcia:2018hlc}.
  \item If $n>p$, it has a maximum located at $x_{max} = p/(2n -
    p)$ \cite{Linder:2015qxa}, whose height for viable models is controlled by
    $n$ and the transition regime slope by $p-n$. Around the inflection points
    the field evolution is fast \cite{Garcia-Garcia:2018hlc}.
\end{itemize}

The observational constraints obtained with Planck 2015 \cite{Ade:2015rim},
BAO from BOSS DR12 \cite{Alam:2016hwk} and supernovae (through an $E(z)$ estimate with
Pantheon compressed sample) \cite{Riess:2017lxs} show that large $\alpha$ and
$\psi_{ini}$ are favored as they make the field move slowly. For the same
reason, $p\sim n$ is also preferred. Finally, despite of having viable
tachyonic solutions, this model would not cause dark energy to cluster
significantly, since dark energy perturbation do not have enough time to grow
appreciably \cite{Garcia-Garcia:2018hlc}.

Note that the lower boundary of $\psi_{ini}$ depends on the maximum value accessible
for $\alpha$, since the total evolution of the field is inversely related to
$\alpha$ (second item of previous list). As a consequence, constraining
$\alpha$ through Eq.~\ref{eq:ns} and $n_s$ and $r$ measurements could
significantly improve the previous result of Ref.
\cite{Garcia-Garcia:2018hlc}, cutting out a big portion of the
available space for $\psi_{ini}$. Quantitatively, 
\begin{equation}
  \alpha = \frac{r}{3 (1 - n_s)^2}\,,
  \label{eq:alpha}
\end{equation}
so that if Stage 4-CMB experiments measured $r<0.01$, $\alpha \lesssim 3.5$,
which would highly constrain the initial position of the field, restricting
its values to positions close to the plateau or the maximum, where the field
slow-rolls. It would be even more dramatic if $r \sim 10^{-3}$ (exhausting the
intended minimum uncertainty \cite{Abazajian:2013vfg}), as $\alpha \sim 0.35$.
However, if $r$ remained close to its upper bound value of Planck 2018 results
($r \sim 0.1$) \cite{Akrami:2018odb}, $\alpha \sim 38$. Then, the available
space for $\psi_{ini}$ would expand towards values closer to the potential
minimum and the inflection points, thanks to the friction $\alpha$ causes.

In this work, we will study the case with $r\lesssim 0.01$, which will allow
to explore the regimes in which the model is both close and different to a
cosmological constant. The mild upper bound in $\alpha$ ($\alpha
  \lesssim 3.5$) will restrict $\psi_{ini}$ to values where the field does
not roll down fast, but far enough to the plateau and maximum, allowing
  a mild evolution. A tighter constraint, as that set by the most precise
  expected measurement of $r$, $r \sim 10^{-3}$~\cite{Abazajian:2013vfg},
  fixes $\alpha \sim 0.35$, pushing the equation of state towards $w \sim
  -1$ to avoid the parts of the potential where the field would move fast and
  yield inviable models, letting alone the maximum (instable), the plateau
  and their closest points. In addition, we choose to avoid the high
  $\alpha$ regime (i.e. high $r$) as it was shown to be unbounded by current
  data \cite{Garcia-Garcia:2018hlc}. The only hope of constraining the
  parameter space relies on finding new data that favors a model sufficiently
  distinct to $\Lambda$CDM and, in that case, our choice of $r\lesssim 0.01$
  ($\alpha \lesssim 3.5$) is broad enough to account for a wide range of
  cosmologies that deviate from $\Lambda$CDM.

\section{Observational probes}
\label{S:probes}
A cohort of next-generation cosmology experiments will collect an unprecedented
amount of data during the next decade, which will allow us to vastly improve our
understanding of cosmology. Our forecasts will include two experiments modelled
after two of the most promising facilities: CMB Stage-4 and the Large Synoptic
Survey Telescope (LSST). The assumptions made to describe these datasets will be
described here. 

\subsection{CMB Stage 4}

Third-generation CMB experiments, such as ACTPol \cite{Calabrese:2014gwa},
SPT-3G \cite{Benson:2014qhw}, BICEP2/Keck \cite{Array:2015xqh} or Simons Array
\cite{Suzuki:2015zzg} will be progressively upgraded to an Stage 4 experiment,
increasing the number of detectors, frequency channels, and sky coverage, allowing
us to cover around $40\%$ of the sky, with a white-noise level $\sigma_T \sim 1\mu
K$-arcmin in temperature \cite{Abazajian:2013vfg}.

S4 will measure primordial CMB temperature and polarization anisotropies as
well as the reconstructed CMB lensing convergence, among other secondary
anisotropies. These measurements will be limited in resolution by the beam size.
We assume a Gaussian beam with a full width at half maximum $\theta_{FWHM}=3$ arcmin.
The corresponding noise power spectrum, assuming white noise, is given by
\begin{equation}
  N_\ell^{x} = \sigma_x^2 \exp\left[\ell(\ell+1) \theta^2_{FWHM}/8\log 2\right]\,,
\end{equation}
where $x$ stands for temperature ($T$) or polarization maps ($E$) and
$\sigma^2_x$ is in units of $\unit{}[\mu K^2 {\rm sr}]$ (we assume $\sigma_E =
\sigma_T \sqrt{2}$). At large scales, statistical and systematic uncertainties,
associated to ground-based facilities such as atmospheric contamination dominate,
and therefore we discard multipoles $l<30$ and use the Planck noise levels in this
regime \cite{Ade:2013zuv} (corresponding to $\sigma_T\simeq30\,\mu{\rm K}$-arcmin.
Furthermore, given the contamination in the temperature power spectrum by
astrophysical foregrounds, we choose different scale cuts for polarization 
($\ell_{\rm max} = 5000$) and temperature ($\ell_{\rm max} = 3000$) multipoles.

Lensing noise is obtained by quadratic combinations of the CMB maps
\cite{Lewis:2006fu} and estimating the reconstruction noise with the minimum
variance weighting, by combinations of the $TT\,,TE\,,TB\,,EB$ and $EE$
individual estimators. This technique significantly reduces the noise of
individual estimators which are noise limited at high-$\ell$ \cite{Hu:2001kj}.
We include CMB lensing information in the range $30 < \ell < 3000$.

\subsection{The Large Synoptic Survey Telescope}

The Large Synoptic Survey Telescope (LSST) is a photometric Stage 4 experiment
that will cover around $20,000\,{\rm deg}^2$ and reach a limit magnitude
$r \sim 27$ \cite{Abell:2009aa}. Photometric catalogs are dense and deep,
which make them excellent for weak lensing studies and multi-probe analyses,
where one does not need high accuracy on the spatial distribution of the
tracers or clustering statistics at small scales. 

Photometric surveys infer the individual galaxies redshifts from their fluxes
in a few broad frequency bands and, as a consequence, have large uncertainties
in the radial clustering pattern. This procedure, will allow LSST to
obtain constraints from different sources: tomographic galaxy clustering and
cosmic shear, galaxy cluster counts, SN Ia and strong lenses.  Among these,
the combination of galaxy clustering and cosmic shear is the most promising
source of information for LSST when combined with measurements of the distance-redshift
relation (through e.g. supernovae or baryon acoustic oscillations). We will follow
Ref. \cite{Alonso:2015uua} in the modelling of both tracers.
\begin{itemize}
  \item \underline{Galaxy clustering}. For galaxy clustering, the most
    relevant observable is the shape of the angular power spectrum or the
    two-point correlation function of the galaxy distribution. In tomographic
    clustering, we divide the galaxy sample in redshift bins and compute the
    auto- and cross-correlation functions between them. In order to simplify
    the analysis, we assume that galaxies can be grouped in two different
    categories -- red galaxies (early-type, elliptical and high-bias) and blue
    galaxies (late-type, spiral and low-bias).  This is just an approximation,
    since red spiral galaxies exist, for example, but it is based on the
    strong bimodality of the galaxy color space \cite{Strateva:2001wt}.
    For instance, red galaxies are less abundant, but show strong
      features in their spectra that allow to extract more accurate photo-$z$
      distributions. Furthermore, they also show a higher clustering amplitude
      (i.e. they have a larger galaxy bias) than their blue counterparts. In
      addition tho these two classes of galaxies, we group together all
    galaxies whose magnitude is above $r\simeq 25.3$, which correspond to the
    so-called `gold sample' of LSST \cite{Abell:2009aa}, and will be used as
    the galaxy shear sample for weak lensing.  In galaxy clustering, the main
    source of statistical noise is shot noise and, following Ref.
    \cite{Alonso:2015uua}, the noise power spectra is given by
    \begin{equation}
      N_l^{ij} = \frac{\delta_{ij}}{n^i}\,
      \label{eq:Nl}
    \end{equation}
    where $n^i$ is the angular number density of galaxies in the $i$-th tomographic
    bin, characterized by its window function $w^i(z)$,
    \begin{equation}
      n^i = \int_{0}^\infty dz\, \bar N(z) w^i(z)\,.
      \label{eq:ni}
    \end{equation}
    We assume, Gaussian photo-$z$ distributions ($p(z_p|z)$), for which the window
    function is
    \begin{eqnarray}
      w^i(z) &=& \int_{z_0^i}^{z_f^i} dz_p\, p(z_p|z) \\ 
       &=& \frac{1}{2} \left[\erf\left(\frac{z - z_0^i}{\sqrt{2}\sigma_z}\right) -
          \erf\left(\frac{z - z_f^i}{\sqrt{2}\sigma_z}\right) \right]\,.
      \end{eqnarray}
    Here $\sigma_z$ is the Gaussian photo-$z$ standard deviation, which we parametrize
    as $\sigma_z(z) = \sigma_0 (1+z)$. We use $\sigma_0 = 0.02$ for red galaxies and
    $\sigma_0 = 0.05$ for the blue and gold samples (red galaxies have usually more
    precise photo-$z$ due to their stronger spectral features). Finally, the
    list of initial and final redshifts for each redshift bin can be found in
    Table~\ref{tab:lsst-bins}, and the galaxy distributions in
    Fig.~\ref{fig:nz}. Note that the redshift spacing was chosen such
      that the width of each bin is equal to 3 times the photo-$z$ uncertainty
      at the center of the bin. This is a compromise between the need to
      sample the redshift range sufficiently well, and avoiding strong
      correlations between different bins due to their overlap.

    \begin{figure}[htb]
      \centering
      \includegraphics[width=0.6\textwidth]{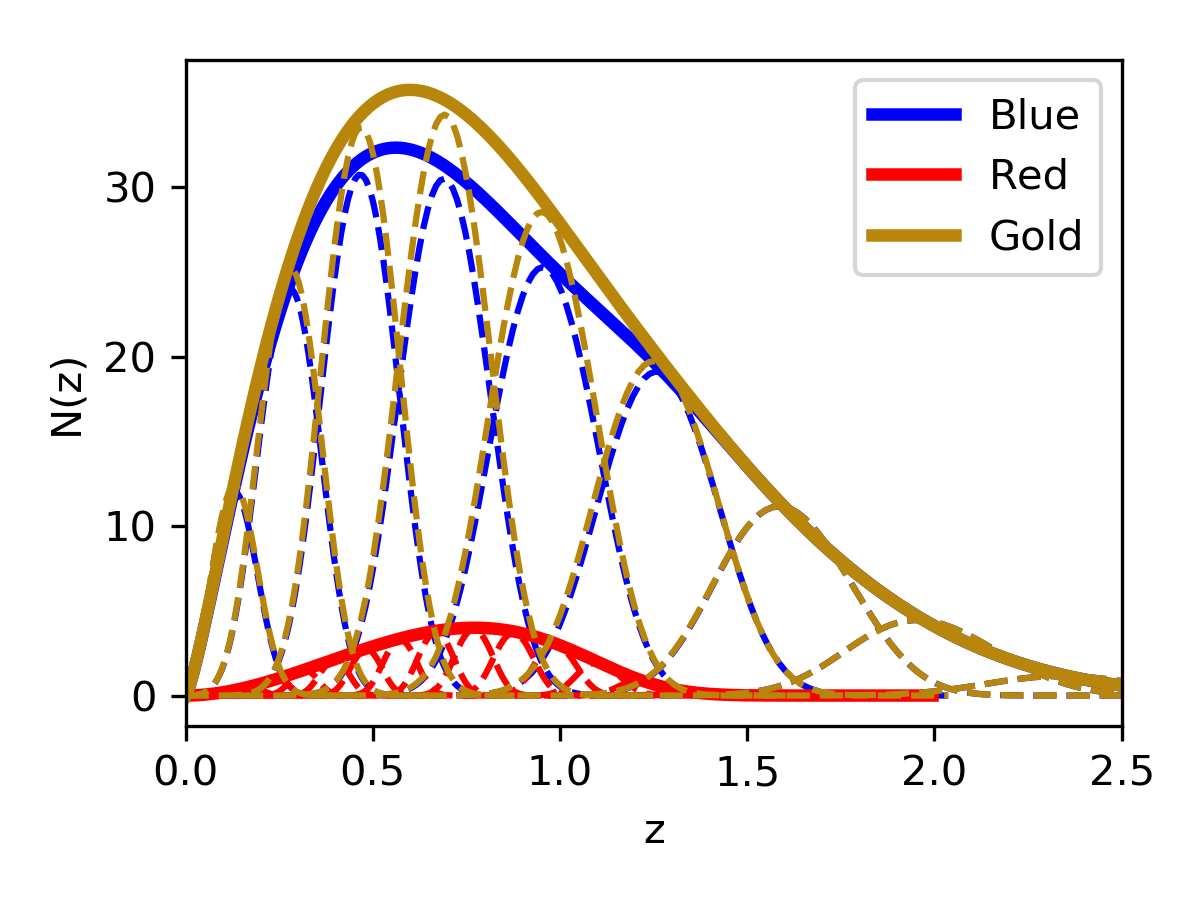}
      \caption{Galaxy density distributions for red, blue and gold samples of
        LSST.  Dashed lines show the windows functions ($W^i \propto \bar N(z)
        w^i(z)$) for each redshift bin.}
      \label{fig:nz}
    \end{figure}

    The main source of uncertainty for galaxy clustering is the relation between
    the galaxy and matter overdensities. On sufficiently large scales, this relation
    is assumed to be linear, and the proportionality constant is the so-called galaxy
    bias $b$ \cite{1996MNRAS.282..347M}. We use a model for the bias of red and blue
    galaxies as
    \begin{equation}
      b_{red}(z) = 1 + z, \qquad b_{blue} = 1 + 0.84z\,.
    \end{equation}
    This is motivated by simulations \cite{Weinberg:2002rm} and observations 
    \cite{Coil:2007jp}, and takes into account the stronger clustering properties
    of red galaxies. 
    
    Given that the linear bias parametrization breaks on small scales, our scale cuts
    for galaxy clustering need to be more conservative. We will define it in a
    redshift dependent manner as $\ell_{\rm max}(\overline z) = \chi(\overline z)
    k_{\rm max}$, where $\overline z$ is the mean redshift of the redshift bin,
    $\chi$ is the radial comoving distance and $k_{\rm max}$ is the threshold
    comoving scale, which we choose to be $k_{\rm max} = \unit[0.2 h]{Mpc^{-1}}$.
    This is the scale up to which a good estimate of the covariance matrix of the
    matter power spectrum in the quasi-linear regime can be made at $z=0$
    \cite{Mohammed:2016sre}.
    \begin{table}
      \centering
      \begin{tabular}{|c|c|}
        \hline
        Sample & Redshift bin edges \\
        \hline
        Blue (cl) & [0, 0.16, 0.35, 0.57, 0.82, 1.12, 1.46, 1.86, 2.33] \\
        \hline
        Red (cl) \&  & [0, 0.06, 0.13, 0.20, 0.27, 0.35, 0.43, \\
        Gold (sh)    &  [0.52, 0.62, 0.72, 0.82, 0.94, 1.1, 1.2, 1.3]\\
        \hline
      \end{tabular}
      \caption{Redshift bin edges for the angular galaxy density distribution
        of each sample. cl $\equiv$ clustering; sh $\equiv$ shear. Refer to
        appendix B.4.2 in Ref~\cite{Alonso:2015uua} for details on the distributions.}
      \label{tab:lsst-bins}
    \end{table}

      As a final remark, we will neglect the effect of magnification bias, given the
      small effect it has on the final constraints \citep{Lorenz:2017iez}.

  \item  \underline{Galaxy shear}. Weak lensing is an unbiased estimator of
    the projected matter perturbations, and is quantified by correlating the projected
    ellipticities of galaxies. The noise power spectrum is directly proportional to the
    variance of the intrinsic galaxy ellipticities, and inversely to the
    angular projected galaxy number density; i.e. $N_l^{ij} =
    \delta_{ij}\sigma_\gamma^2/n^i$. Here,
    $\sigma_\gamma$ includes both the dispersion of the intrinsic galaxy
    ellipticities and the measurements uncertainties, and is set to
    $\sigma_\gamma = 0.28$ \cite{Abell:2009aa}. We marginalize over shape measurement
    systematics in the form of a free multiplicative bias parameter for each reshift
    bin. Other sources of systematic uncertainty, such as intrinsic
    alignment, shape-measurement systematics or baryonic uncertainties will be
    neglected. We expect their effect on the constraints on the parameter
    space of the $\alpha$-attractor dark energy model to be negligible compared to
    other sources of systematic uncertainties, particularly the multiplicative bias.
    We will however impose a scale cut $\ell\leq 2000$ to avoid uncertainties
    associated with the modelling of baryonic effects in the matter power
    spectrum \cite{Rudd:2007zx,vanDaalen:2011xb,
      vanDaalen:2013ita,Hellwing:2016ucy,Harnois-Deraps:2014sva,Chisari:2018prw,Huang:2018wpy}.
    The redshift bins for the gold sample used for weak lensing are given in
    Table~\ref{tab:lsst-bins}
\end{itemize}

\begin{table}
  \centering
  \begin{tabular}{|l|c|}
    \hline
    Tracer & Noise contribution \\
    \hline
    S4CMB & $N_l^{x} = \sigma_x^2 \exp\left[l(l+1) \theta^2_{FWHM}/8\log 2\right]$\,, \\
    gal.\ cl. & $N_l^{ij} = \delta_{ij}/n^i n^j$\\
    gal.\ sh. & $N_l^i = \sigma_\gamma^2/N_\Omega^i$\\
    \hline
  \end{tabular}
  \caption{Noise contribution to the power spectra for LSST measurements. For
    S4CMB experiments, the noise level $\sigma_x = \sigma_T,\, \sigma_E$,
    where $\sigma_E/\sqrt{2} = \sigma_T \sim 1\mu K$-arcmin and
    $\Theta_{FWHM} = \unit{3}[arcmin]$. For galaxy clustering
    (gal.\ cl.), $\delta_{ij}$ is the identity matrix and $n^i$ is the
    galaxy number in the z-bin $i$, given by Eq.~\ref{eq:ni}. For
    galaxy shear (gal.\ sh.), the variance of the intrinsic galaxy
    ellipticities, $\sigma_{\gamma} = 0.28$ and $N^i_\Omega$ is the angular
    galaxy number density of z-bin $i$.}
  \label{tab:LSST}
\end{table}

\subsection{Spectroscopic Surveys: DESI and WFIRST}

Spectroscopic surveys are especially aimed to study phenomena at smaller
scales, like BAO and redshift-space distortions. The high redshift resolution of
spectroscopic surveys makes a tomographic analysis as described in the previous
section computationally intractable and inefficient. The standard analysis
studying the multipoles of the 3D galaxy power spectrum is however not easy to
incorporate into our forecasting formalism, in terms of fully characterizing the
correlations with overlapping tomographic data.

Instead, we will directly incorporate the BAO forecasts for DESI \cite{Aghamousa:2016zmz} and
WFIRST \cite{Spergel:2013tha,Font-Ribera:2013rwa}, using the error estimates summarized in
Table~\ref{tab:BAO}. The errors are given on the angular diameter distance ($D_A = (1+z)^{-1} \int_0^z dz' H^{-1}(z')$)
and Hubble parameter ($H(z)$).

\begin{table}
  \centering
\begin{tabular}{|lcc||lcc|}
\hline
\multicolumn{6}{|c|}{BAO error predictions}\\ 
\hline
\multicolumn{3}{|c||}{DESI} & \multicolumn{3}{|c|}{WFIRST}\\
\hline
$z$ & $ \frac{\sigma_{D_A/s}}{D_A/s}(\%)$ & $\frac{\sigma_{Hs}}{Hs}(\%)$ &
$z$ & $ \frac{\sigma_{D_A/s}}{D_A/s}(\%)$ & $\frac{\sigma_{Hs}}{Hs}(\%)$ \\
\hline
0.05 & 6.12 & 12.10 & 1.05 & 1.51 & 2.72 \\ 
0.15 & 2.35 & 4.66  & 1.15 & 1.43 & 2.56 \\ 
0.25 & 1.51 & 2.97  & 1.25 & 1.35 & 2.42 \\ 
0.35 & 1.32 & 2.44  & 1.35 & 1.29 & 2.30 \\ 
0.45 & 2.39 & 3.69  & 1.45 & 1.24 & 2.21 \\ 
0.65 & 0.82 & 1.50  & 1.55 & 1.23 & 2.16 \\ 
0.75 & 0.69 & 1.27  & 1.65 & 1.25 & 2.15 \\ 
0.85 & 0.69 & 1.22  & 1.75 & 1.28 & 2.16 \\ 
0.95 & 0.73 & 1.22  & 1.85 & 1.33 & 2.19 \\ 
1.05 & 0.89 & 1.37  & 1.95 & 1.41 & 2.27 \\ 
1.15 & 0.94 & 1.39  & 2.05 & 2.51 & 3.52 \\ 
1.25 & 0.96 & 1.39  & 2.15 & 2.60 & 3.62 \\ 
1.35 & 1.50 & 2.02  & 2.25 & 2.74 & 3.78 \\ 
1.45 & 1.59 & 2.13  & 2.35 & 3.02 & 4.09 \\ 
1.55 & 1.90 & 2.52  & 2.45 & 3.38 & 4.52 \\ 
1.65 & 2.88 & 3.80  & 2.55 & 3.87 & 5.11 \\ 
1.75 & 4.64 & 6.30  & 2.65 & 4.52 & 5.90 \\ 
1.85 & 4.71 & 6.39  & 2.75 & 5.41 & 6.99 \\ 
\hline
\end{tabular}
\caption{ WFIRST and DESI BAO errors. Respectively, they have been taken from 
  Table VII in Ref.~\cite{Font-Ribera:2013rwa} and Tables 2.3 and 2.5 of
  Ref.~\cite{Aghamousa:2016zmz}. The early-time BAO error predictions from
  Ly-$\alpha$ and quasars (QSO) have been omitted as they are above the $H$ and
  $D_A$ partial derivatives (see Fig.~\ref{fig:derivatives}) values, having
  little contribution to the total Fisher matrix.}
\label{tab:BAO}
\end{table}

DESI \cite{Aghamousa:2016zmz} will cover $\sim \unit[14000]{deg^2}$ from the
North Hemisphere and target Luminous Red Galaxies (LRGs), Emission Line
Galaxies (ELGs) and quasars. Their BAO $D_A$ and $H$ error estimations cover
18 redshifts, uniformly distributed in the redshift range $0.05 \leq z \leq
1.85$. The details of their forecast analysis can be found in
Ref.~\cite{Aghamousa:2016zmz}. 
%
On the other hand, WFIRST will measure redshifts for $\sim 2.6 \times 10^7$
galaxies over $\sim \unit[2000]{deg^2}$. Their forecast assumed the galaxy
number densities from from Ref.~\cite{Spergel:2013tha}. 
We use forecast errors on the BAO parameters over 18 redshift bins, uniformly
distributed in $1.05 \leq z \leq 2.75$. Additionally, WFIRST will also be
able to measure the expansion of the Universe through type Ia supernovae. We will
include this probe through the forecast for $E(z) = H(z)/H(0)$
from Ref.~\cite{Riess:2017lxs} (see Table~\ref{tab:Ez}), the same way this
was done in Ref.~\cite{Garcia-Garcia:2018hlc}. We will neglect correlations
between different redshifts as this effect is negligible in comparison with the
constraining power of the other experiments. The predictions for $E(z)$ were obtained
from a simulation based on Ref.~\cite{Hounsell:2017ejq}, plus an external sample at
$z < 0.1$. The predictions for $E(z)$ for WFIRST are based in simulations done
by  Ref.~\cite{Hounsell:2017ejq} for WFIRST, where the systematic errors in
the adopted model fall below the statistical errors. The number of supernovae
in each redshift bin is shown in Table~\ref{tab:SN}.

\begin{table}
  \centering
  \begin{tabular}{|lc|}
    \hline
    \multicolumn{2}{|c|}{WFIRST SN Ia}\\
    \multicolumn{2}{|c|}{$E(z)$ predictions}\\
    \hline
    $z$ & $\sigma(E(z)) (\%)$\\
    \hline
    0.07 & 1.3\\
    0.2  & 1.1\\
    0.35 & 1.5\\
    0.6  & 1.5\\
    0.8  & 2.0\\
    1.0  & 2.3\\
    1.3  & 2.6\\
    1.7  & 3.4\\
    2.5  & 8.9\\
    \hline
  \end{tabular}
  \caption{$E(z) = H(z)/H(0)$ estimated relative errors from WFIRST SN Ia.
    Values have been taken from Table 7 of Ref.~\cite{Riess:2017lxs}, based
    on Ref.~\cite{Hounsell:2017ejq} simulations plus an external sample at
    $z < 0.1$.}
  \label{tab:Ez}
\end{table}

\begin{table}[htb]
  \centering
  \begin{tabular}{|c|c|}
    \hline
    $z \in$    & SN \\
    \hline
    $[0, 0.1]  $ & 800 \\
    $[0.1, 0.4]$ & 557 \\
    $[0.4, 0.8]$ & 4807 \\
    $[0.8, 1.7]$ & 5892 \\
    \hline
  \end{tabular}
  \caption{Redshift bins and number of supernovae obtained in a realistic
    simulation of the Imaging All-z observational strategy
    \cite{Hounsell:2017ejq}, used in Ref~\cite{Riess:2017lxs} to forecast the
    uncertainties on $E(z)$. The first 800 SN are assumed to be obtained from a
    different experiment.}
  \label{tab:SN}
\end{table}

\section{Fisher formalism}
\label{S:Fisher}
This section  summarizes the Fisher formalism introduced in
Ref. \cite{Alonso:2015sfa}. Each projected probe (CMB primary and
lensing, photometric galaxy clustering and cosmic shear) labelled as $a$
is composed of a number of sky maps $N^a_{\rm maps}$, which can be fully
described by their harmonic coefficients ($a_{\ell m}^{a\,, i}$, $i \in [0,
N^{a}_{\rm maps}]$). They can be grouped into a vector $\bfa_{\ell m}$, the
covariance matrix of which is the power spectrum::
\begin{equation}
  \langle \bfa_{\ell m} \bfa_{\ell' m'}^\dag \rangle = \delta_{\ell\ell'} \delta_{mm'} {\sf C}_l\,.
\end{equation}
Under the assumption of being Gaussian distributed, the likelihood is given by
\begin{equation}
  -2\log L = \sum_{\ell_{\rm min}}^{\ell_{\rm max}} \sum_{m=-\ell}^{m=\ell} 
  \left[\bfa^\dag_{\ell m} {\sf C}_\ell^{-1} \bfa_{\ell m} + \log(\det(2\pi {\sf C}_\ell)) \right]\,,
\end{equation}
which can be expanded around the maximum in order to find the Fisher matrix
\begin{equation}
  F_{\mu\nu} = \sum_{\ell=\ell_{\rm min}}^{\ell_{\rm max}}
  f_{\rm sky} \frac{2\ell + 1}{2} \tr\left(\partial_\mu C_\ell C_\ell^{-1} \partial_\nu C_\ell C_\ell^{-1}\right).
\end{equation}
The covariance matrix of the parameters $\theta$ can then be obtained by inverting $F$. In the previous
equation, $\partial_\mu$ is the partial derivative respect to the parameter $\theta_\mu$ and $f_{sky}$ is
the sky fraction covered by the considered probes.

Furthermore, we will assume that noise and cosmological signal are uncorrelated in the observed anisotropies,
i.e. given $\bfa_{\ell m} = \bfs_{\ell m} + \bfn_{\ell m}$, ${\sf C}_\ell = {\sf C}_\ell^S + {\sf C}_\ell^N = \langle \bfs_{\ell m}\bfs_{\ell m}^\dag \rangle +
\langle \bfn_{\ell m}\bfn_{\ell m}^\dag\rangle$, where ${\sf C}^S_\ell$ and ${\sf C}^N_\ell$ are the signal and 
noise power spectra.

The Fisher matrix with DESI and WFIRST probes will be computed as
\begin{equation}
  F_{\mu\nu} = \sum_i \frac{\partial_\mu q_i\partial_\nu q_i}{\sigma_i^2},
\end{equation}
where $q_i$ is the measurement of a given quantity $q$ (which stands for $D_A(z)$,
or $E(z)$) in the $i$-th redshift bin, and $\sigma_i^2$ is the forecasted error
on that quantity. This Fisher matrix is added to the one computed for CMB and 
photometric survey data. This ignores possible correlations between both sets of
observables. We do not expect our results to be very sensitive to this assumption.

Finally, all partial derivatives with respect to $\theta_\mu$ will be computed via finite central
differences,
\begin{equation}
  \partial_\mu f= \frac{f(\theta_\mu + \delta\theta_\mu) - f(\theta_\mu -
  \delta\theta_\mu)}{2\delta\theta_\mu} + O(\delta \theta^3).
\end{equation}
In addition, the power spectra, ${\sf C}_\ell^S$, will be obtained using
{\tt hi\_class} \cite{Zumalacarregui:2016pph}, a modified version of CLASS
\cite{Blas:2011rf} that incorporates Horndeski models \cite{Horndeski:1974wa}
without assuming the quasi-static approximation, which
ensures results are valid at scales larger than the sound horizon
\cite{Sawicki:2015zya}. Finally, we will use the Limber approximation
\cite{1953ApJ...117..134L} in the full range of scales. The software used
to combine all these ingredients is available
online\footnote{\url{https://gitlab.com/ardok-m/GoFish_aatt-forecast/tree/aatt},
  a modified version of \url{https://github.com/damonge/GoFish} (the
  master branch of the former repository)}.

\section{Results}
\label{S:results}

The next generation of data, despite its increase on accuracy, will fail to fully
constrain the generalized $\alpha$-attractor model, as present observations
did, if they continue preferring a $\Lambda$CDM background evolution. One must
recall that this model has an infinite region of the parameter space that is
indistinguishable from $\Lambda$CDM, corresponding to large $\alpha$ (acts as
a friction to the field motion) or having $\psi_{ini}$ on the plateau (or,
with more fine tunning, close to a maximum) \cite{Garcia-Garcia:2018hlc}. 

We will investigate the parameter space that lies $1 \sigma$ off the best fit
result of Ref.~\cite{Garcia-Garcia:2018hlc}, which is able to differentiate
from a cosmological constant. Current observations prefer the 
  cosmological constant-like regime, which correspond to an unbounded region
  on the parameter space~\cite{Garcia-Garcia:2018hlc}. Therefore,
  as we said before, if future observations were to continue favoring
  $\Lambda$CDM, they would not be able to constrain the parameter space. At
  best, they will be able to rise the lower bounds for $\alpha$ and
  $\psi_{ini}$. As a consequence, the only hope of finding tight constraints
  relies on new data that favors models slightly different (they still have to
  be compatible with current observations, at their level of accuracy) to
  $\Lambda$CDM. This regime correspond to the parameters $1\sigma$ off the
  best-fit of Ref.~\cite{Garcia-Garcia:2018hlc}.

\begin{figure}[htb]
  \centering
  \includegraphics[width=\textwidth]{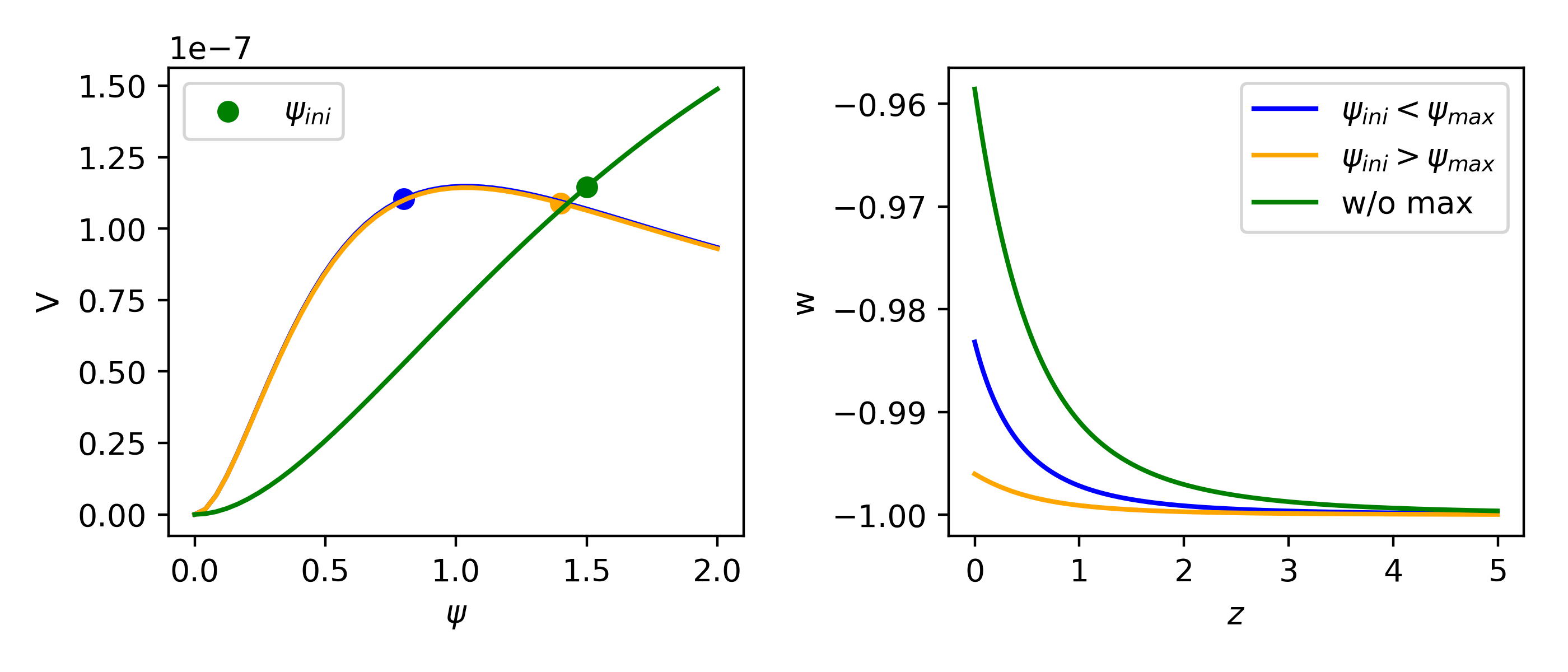}
  \caption{Fiducial potentials ($V$) and equations of state ($w$). The dots
    mark the initial position of the field. }
  \label{fig:V-w}
\end{figure}

The cosmological parameters have been chosen as
in Table 3 of Ref.~\cite{Garcia-Garcia:2018hlc}, i.e. $\Omega_{cdm}h^2 =
0.1183$, $\Omega_{b}h^2 = 0.02221$, $h = 0.682$, $10^9 A_s = 2.14$, $n_s =
0.9649$, $\tau_{reio} = $ = 0.067. For the $\alpha$-attractor parameters, we study two distinct
cases, corresponding to models with and without a maximum ($p \geq n$ and $p < n$, respectively).
In the first case, the $1\sigma$-off parameters are $\{\psi_{ini},\, \alpha,\,
p,\, n\} = \{1.5,\, 2,\, 2,\, 1\}$. In the second case, we choose $\{\alpha, p, n\} = \{2, 2,
3.5\}$ and we study two further options for $\psi_{ini}$: $0.8$ and $1.4$. This corresponds
to the cases with $\psi_{ini}$ smaller and larger than $\psi_{max} = 1.04$
(see Fig.~\ref{fig:V-w}). A summary of the fiducial models and constraints can
be found in Table~\ref{tab:results}. In next sections, we will discuss them in
detail.  The $c$ parameter, which fixes the potential amplitude, is fixed via
the Friedman equation ($ 1 = \sum_i \Omega_i$).


\begin{table}
  \centering
  \begin{tabular}{|c||c|c|c|c||c|c|}
    \hline
    Pars.          & $\psi_{ini}$ & $\alpha$ & $p$ & $n$ & FoM & $\mbox{FoM}_{\mbox{\scriptsize CPL}}$\\
    \hline
    \hline
    Fid (no max).  & 1.5 & 2 & 2 & 1 & $-$ & $-$\\
    $\sigma$       & $\lesssim 1$ & 0.5 & $\lesssim 1$ & $\lesssim 1$ & $\sim 10$ & $10^3$\\
    \hline
    \hline
    Fid ($\psi_{ini} < \psi_{max}$).  & 0.8 & 2 & 2 & 3.5 & $-$ & $-$\\
    $\sigma$                          & 0.2 & $\sim 0.5$ & 0.4 & 1 & $\sim 100$ & $10^2$\\
    \hline
    \hline
    Fid ($\psi_{ini} > \psi_{max}$).  & 1.4 & 2 & 2 & 3.5 & $-$ & $-$\\
    $\sigma$                          & 2 & $\sim 1$ & 2 & 2 & $\sim 1 $ & $10^4$\\
    \hline
  \end{tabular}
  \caption{Fiducial values and predicted constraints using all probes. Recall
    that $\psi_{max} = 1.04$. The results with a '$\sim$' sign mean that
    little changes on the numerical derivative yield changes on the first
    significant digit. This is consequence of the strong correlations between
    the model parameters. $\mbox{FoM} \equiv \mbox{FoM}(\psi_{ini},\,
    \alpha,\, p,\, n)$ and $\mbox{FoM}_{\mbox{CPL}} \equiv \mbox{FoM}(w_0,\,
    w_a)$. The potential shape when $\psi_{ini} < \psi_{max}$ allows for a
    larger variety of evolutions and, therefore, of $w_0 - w_a$ values. This
    causes a greater $\mbox{FoM}_{\mbox{CPL}}$ than in the other cases.}
  \label{tab:results}
\end{table}

\subsection{Case without maximum: $p > n$}
The fiducial model parameters are $\{\psi_{ini},\, \alpha,\, p,\, n\} =
\{1.5,\, 2,\, 2,\, 1\}$. In Tab.~\ref{tab:FoM} the Figures of Merit (FoM) for different
combinations of the experiments are shown. Recall that the FoM is defined by
\cite{Albrecht:2006um}
\begin{equation}
  \mbox{FoM} = (\det{C})^{-1/2}.
  \label{eq:FoM}
\end{equation}
In our case, the covariance matrix ($C$) is obtained inverting the full Fisher
matrix and marginalizing over the nuisance and cosmological parameters, so
that we describe just the constraining power of the next generation
experiments on the parameter space of the $\alpha$-attractor model. 


\begin{table}
  \centering
  \begin{tabular}{|c|c|c|c|}
    \hline
    & \multicolumn{3}{c|}{FoM($\psi_{ini},\, \alpha,\, p,\, n$)} \\
    \hline
    Experiments & w/o max. & $\psi_{ini} < \psi_{max}$ & $\psi_{ini} > \psi_{max}$\\
    \hline
     SN Ia, BAO, gal.\ sh                       & --                 & -- & --\\ 
     S4CMB                                      & --                 & $2 \times 10^{-1}$  & --\\
     S4CMB + BAO                                & $6 \times 10^{-2}$ & $7 \times 10^{-1}$  & -- \\
     S4CMB + SN Ia                              & $1 \times 10^{-1}$ & $1$                 & --\\
     {\footnotesize S4CMB + BAO  + SN Ia }      & $1 \times 10^{-1}$ & $2$                 & --\\
     gal.\ cl                                   & $2 \times 10$      & $1 \times 10^{2}$   & $3 \times 10^{-1}$\\
     {\footnotesize S4CMB + gal.$^{*}$ + SNIa} & $4 \times 10$      & $3 \times 10^{2}$   & $2$\\
     All                                        & $4 \times 10$      & $3 \times 10^{2}$   & $2$\\
    \hline
  \end{tabular}
  \caption{Figures of Merit for different combinations of future experiment
    measurements on the parameter space of the $\alpha$-attractor model. BAO
    combines DESI and WFIRST predictions, SN Ia comes from constraints on
    $E(z)$ using WFIRST forecasts \cite{Riess:2017lxs} and galaxy clustering
    (gal.\ cl) and shear (gal.\ sh) are those from LSST. The combination of
    galaxy clustering and shear has been written as gal$^{*}$. We have omitted
    FoM$ < 10^{-1}$, considering those as unable to constraint the parameter
    space.}
  \label{tab:FoM}
\end{table}

\begin{figure}[tbh]
  \centering
  \includegraphics[width=1\textwidth]{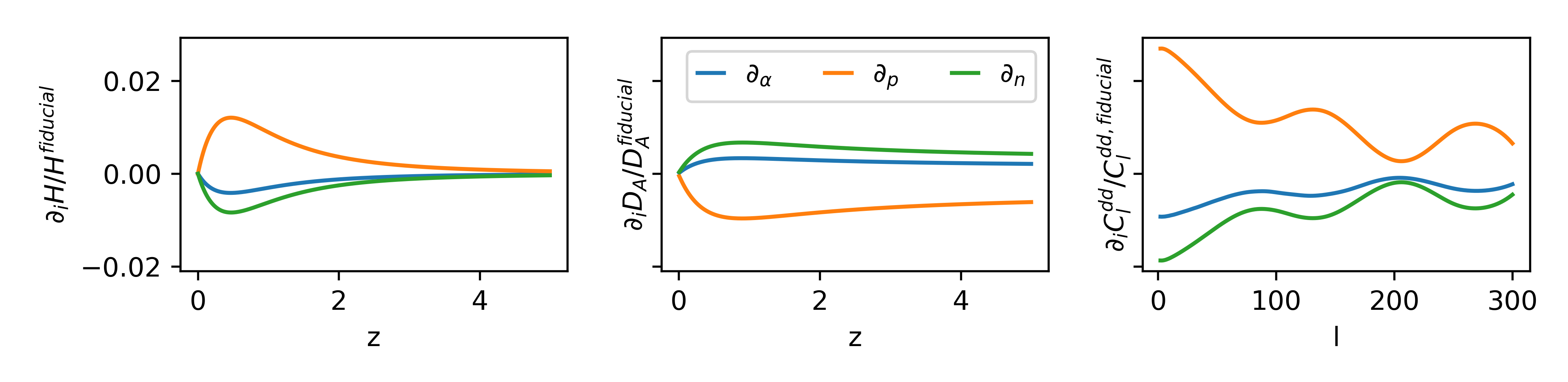}
  \caption{Numerical derivatives of the Hubble parameter, angular diameter
    distance and matter correlation. The computation was done for the case
    without maximum.}
  \label{fig:derivatives}
\end{figure}
\begin{figure}[tbh]
  \centering
  \includegraphics[width=0.6\textwidth]{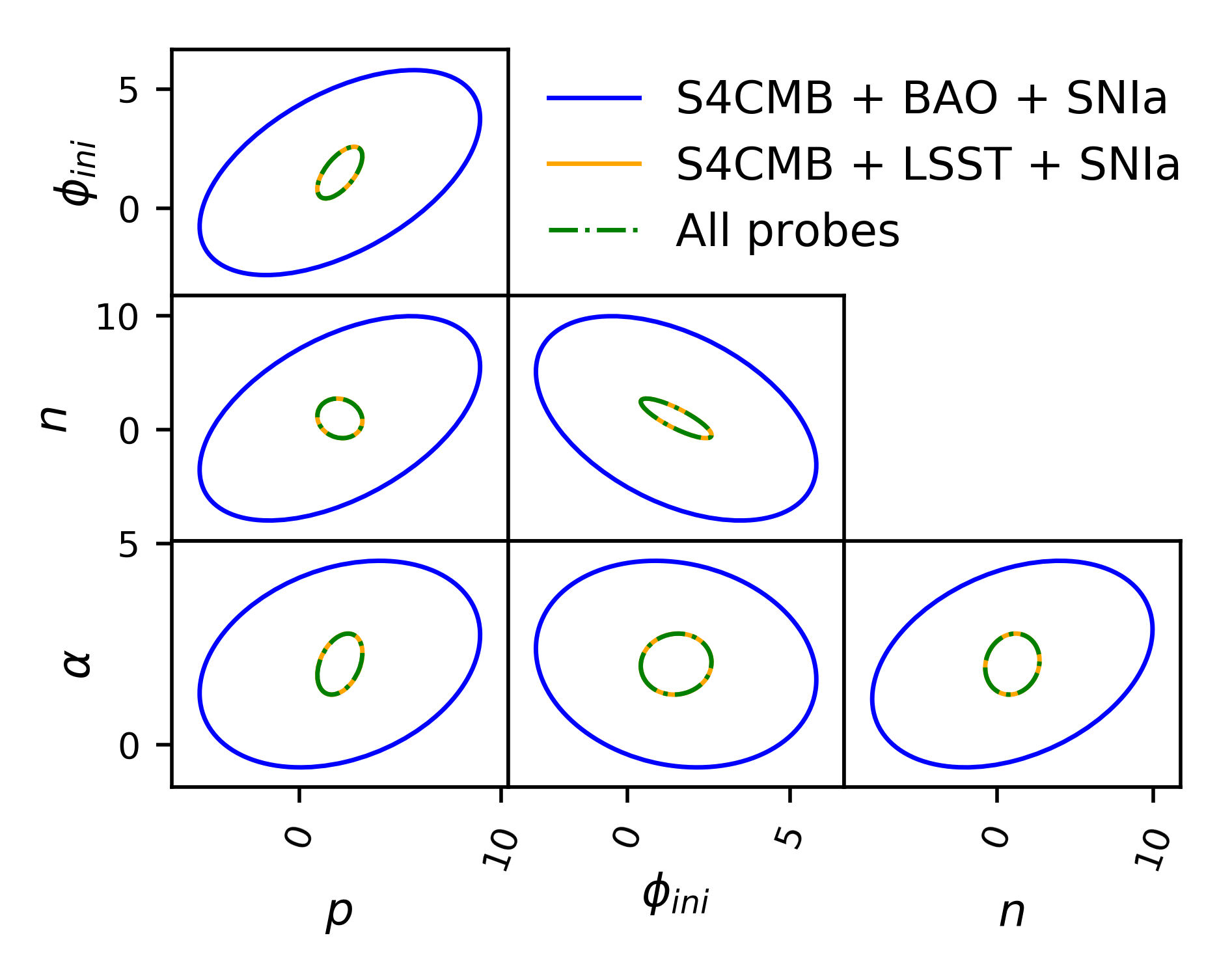}%
  \caption{$2\sigma$-regions for the model parameters when the potential has
    no maximum. Note that the maximum constraints are already found for S4CMB
    + LSST + SNIa.}
  \label{fig:params1}
\end{figure}

Table~\ref{tab:FoM} shows that LSST galaxy clustering is necessary to be able
to constrain the parameter space of the dark energy $\alpha$-attractor model.
The galaxy power spectrum is the observable that is most sensitive to changes
on the model parameters, as shown in Fig.~\ref{fig:derivatives}. Furthermore,
the combination of galaxy clustering and the other probes is able to increase
the FoM almost by a factor 2; exhausting the constraining power of future
observations. In Fig.~\ref{fig:params1} we show the predicted
2$\sigma$-regions for the cases with all probes, with S4-CMB + BAO + SNIa and
S4-CMB + LSST + SNIa. Galaxy clustering would be able to alleviate the
degeneracy between $\alpha$ and $\psi_{ini}$ that made it difficult to find
good constraints in Ref. \cite{Garcia-Garcia:2018hlc}.  The strong degeneracy
between $\psi_{ini}$ and the exponents is such that slight variations of one
can be compensated with any of the other in order to prevent the field from
rolling down the potential too fast. 


\begin{figure}[tbh]
  \centering
  \includegraphics[width=0.6\textwidth]{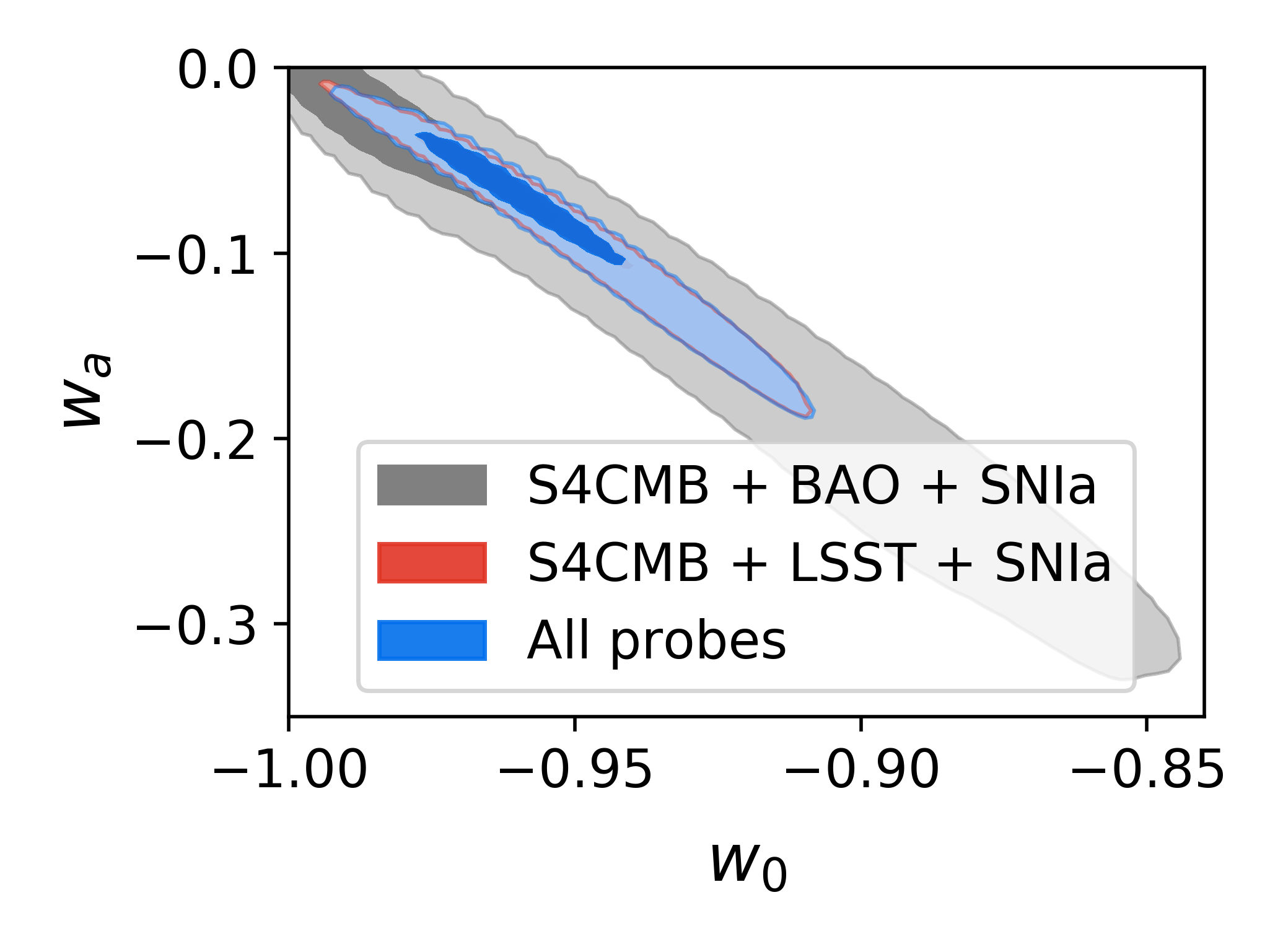}%
  \caption{$2\sigma$-regions of the $w_0-w_a$ parameters parting from a
    fiducial model 1$\sigma$-off the $\Lambda$CDM regime and no maximum. Note
    that the maximum constraints are already found for S4CMB + LSST + SNIa.}
  \label{fig:w0wa1}
\end{figure}

These constraints in the parameter space can also be seen in the CPL
parametrization of dark energy
($w_0-w_a$)~\cite{Chevallier:2000qy,Linder:2002et}. In fact, one can see a
similar improvement on the FoM (Table~\ref{tab:FoM_w0wa}) of $w_0-w_a$ as in
the model parameters (Table~\ref{tab:FoM}). The FoM for the CPL parameters has
been defines as
\begin{equation}
  \mbox{FoM}(w_0, w_a) = \frac{1}{\mbox{area}_{1\sigma}(w_0, w_a)},
  \label{eq:FoM_w0wa}
\end{equation}
which generalizes Eq.~\ref{eq:FoM} for a non elliptical shape.
It must be noted, however, that the main reason behind the large FoM is due to
the fact that this model belongs to the thawing quintessence class, which is
known to have little freedom in the $w_0-w_a$ plane \cite{Linder:2015zxa}.
Interestingly, it could be expected to detect a $2\sigma$ deviation from a
cosmological constant ($w_0 = -1$, $w_a = 0$), provided that the fiducial
model were the true one and one included LSST galaxy clustering observations
(Fig.~\ref{fig:w0wa1}). This would not be the case if galaxy clustering were
not taken into account. In fact, the weak constraints from the other
cosmological probes would shift the $w_0-w_a$ $1\sigma$ region towards $w\sim
-1$.


\begin{table}
  \centering
  \begin{tabular}{|c|c|c|c|}
    \hline
    & \multicolumn{3}{c|}{FoM($w_0,\, w_a$)} \\
    \hline
    Experiments & w/o max. & $\psi_{ini} < \psi_{max}$ & $\psi_{ini} > \psi_{max}$\\
    \hline
    {\footnotesize S4CMB + BAO  + SN Ia}          & $6 \times 10^2$   & $5 \times 10^1$ & $6 \times 10^2$    \\
    {\footnotesize S4CMB + gal.$^*$  + SNIa}      & $2 \times 10^3$   & $2 \times 10^2$ & $8 \times 10^3$    \\
    All                                           & $2 \times 10^3$   & $2 \times 10^2$ & $8 \times 10^3$    \\
    \hline
  \end{tabular}
  \caption{Figures of Merit (Eq.~\ref{eq:FoM}) for different combinations of
    future experiment on the $w_0$-$w_a$ parameters. BAO combines DESI and
    WFIRST predictions, SN Ia comes from constraints on $E(z)$ using WFIRST
    forecasts \cite{Riess:2017lxs} and galaxy clustering (gal.\ cl) and shear
    (gal.\ sh) are those from LSST. The combination of galaxy clustering and
    shear has been written as gal.$^*$.}
  \label{tab:FoM_w0wa}
\end{table}

The $w_0$-$w_a$ contours were obtained: first, we diagonalize the covariance matrix
(i.e. $F^{-1}$). We then take samples of the uncorrelated Gaussian distribution and
transform-back to the original basis. In doing so, we reject any model with $p < n$
(i.e. models with maximum) and with negative model parameters. Once selected,
we used {\tt hi\_class} \cite{Zumalacarregui:2016pph} to compute the
corresponding $w_0$-$w_a$ parameters, with $w_a$ computed as $w_a =
-dw/d\ln(a)|_{a=1}$). Finally, we used {\tt GetDist}
\footnote{\url{https://github.com/cmbant/getdist}} to produce contours of the
corresponding samples. 


\subsection{Case with maximum: $p<n$}

\begin{figure}[tb]
  \centering
  \subfloat[$\psi_{ini} < \psi_{max}$]{%
      \includegraphics[width=0.45\textwidth]{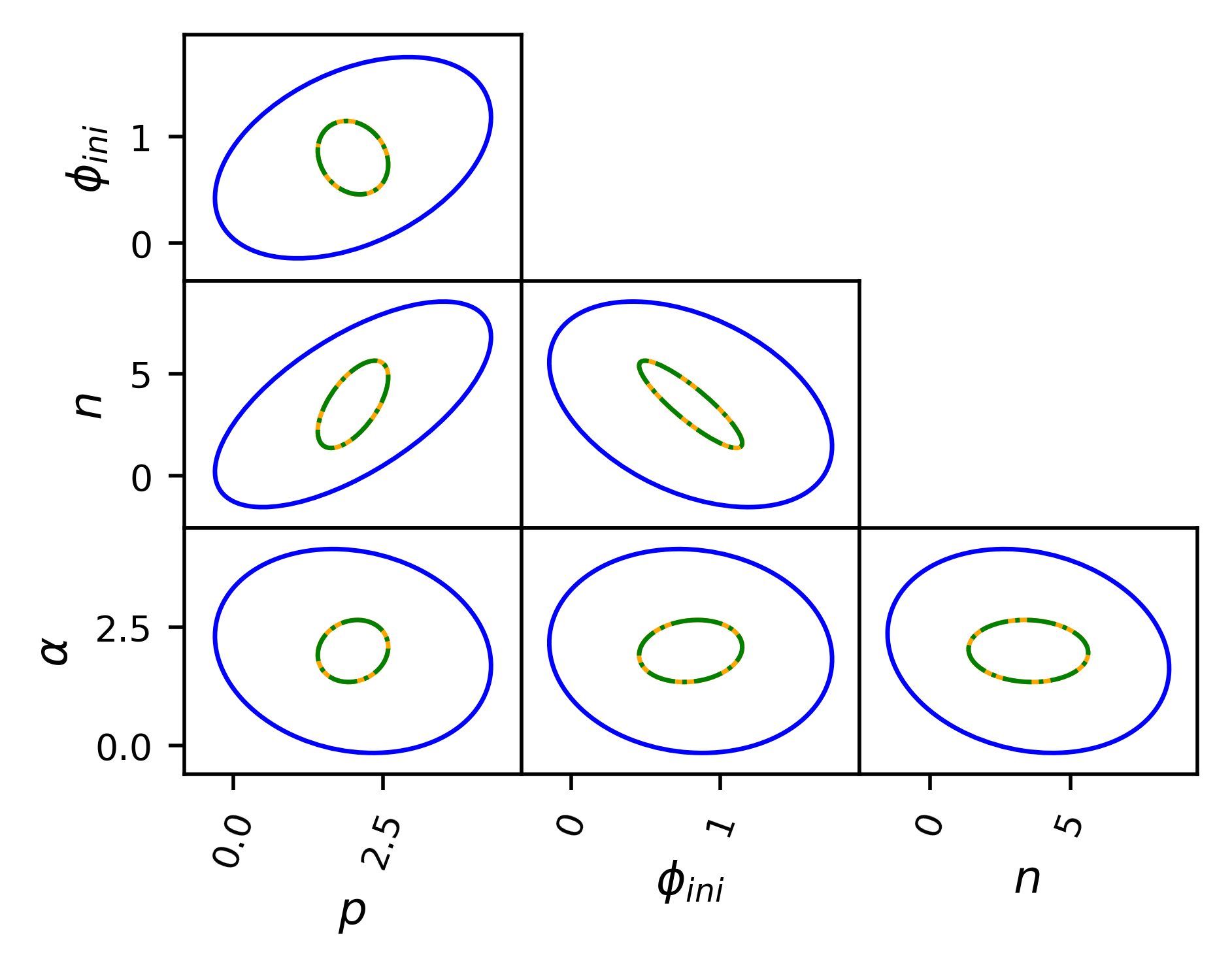}%
  }
  \subfloat[$\psi_{ini} > \psi_{max}$]{%
      \includegraphics[width=0.45\textwidth]{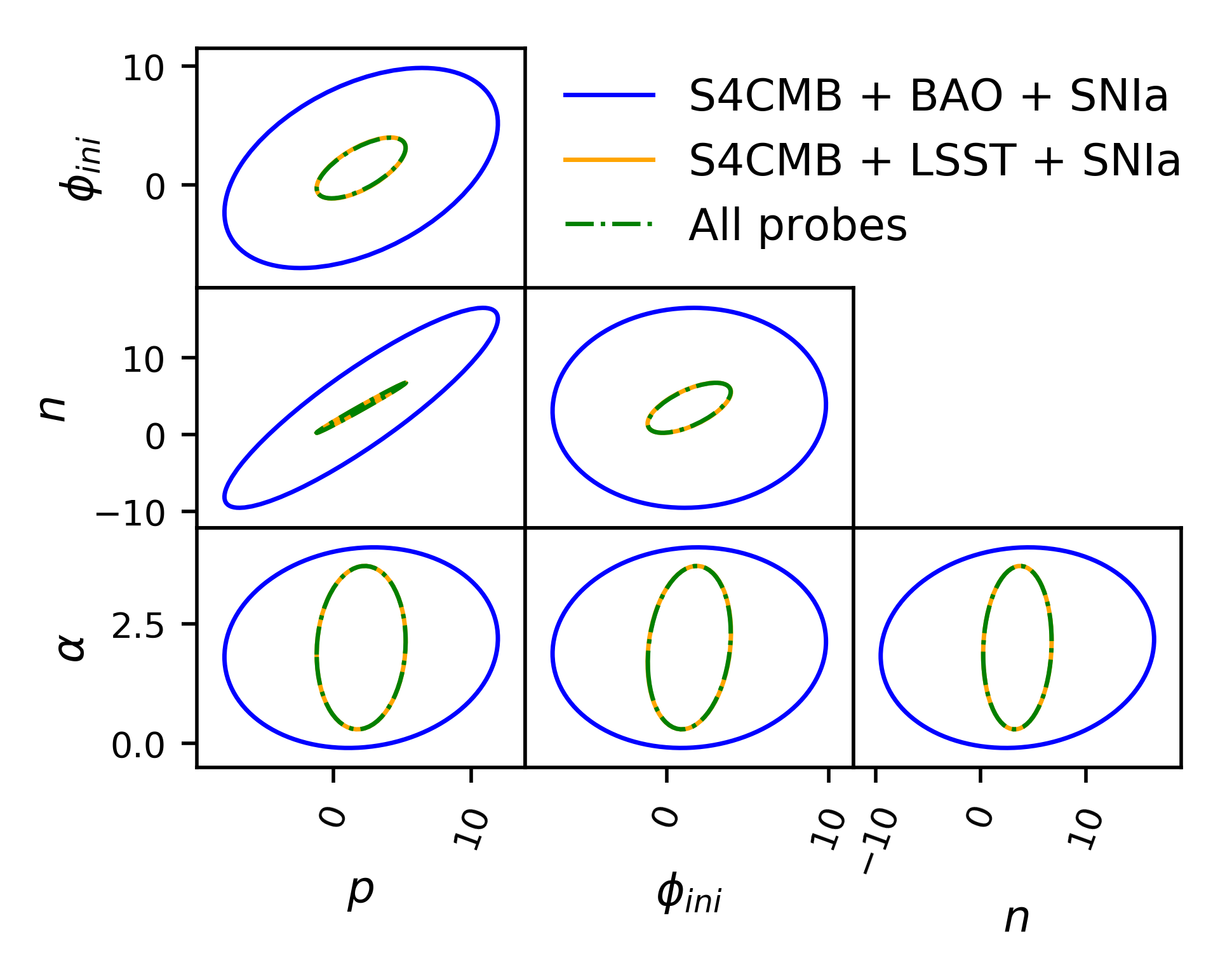}%
  }
  \caption{$2\sigma$-regions for the model parameters when the potential has a
    maximum. Note that the maximum constraints are already found for S4CMB
    + LSST + SNIa.}
  \label{fig:params2}
\end{figure}

\begin{figure}[tb]
  \centering
  \subfloat[$\psi_{ini} < \psi_{max}$]{%
    \includegraphics[width=0.45\textwidth]{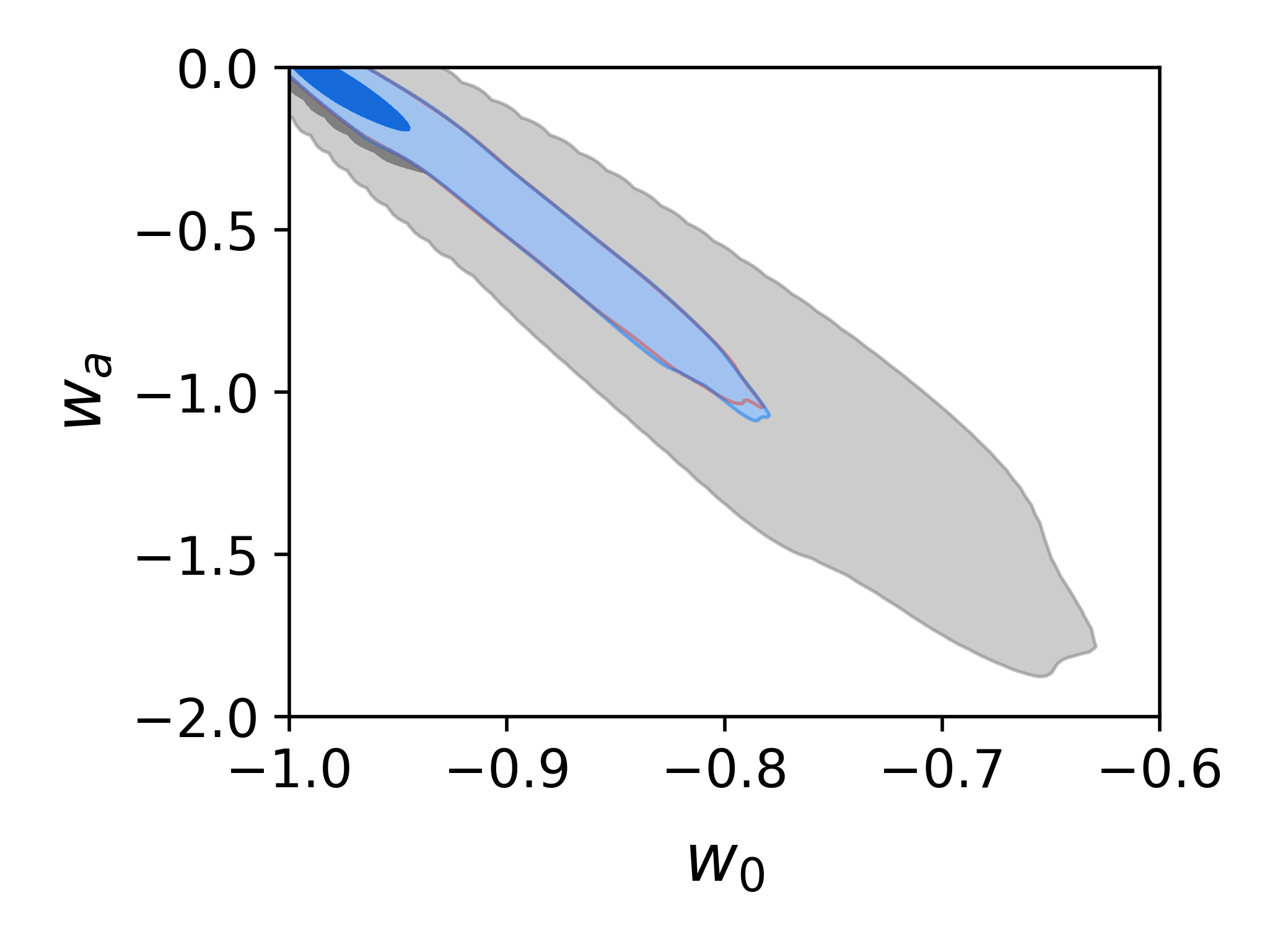}%
  }
  \subfloat[$\psi_{ini} > \psi_{max}$]{%
    \includegraphics[width=0.45\textwidth]{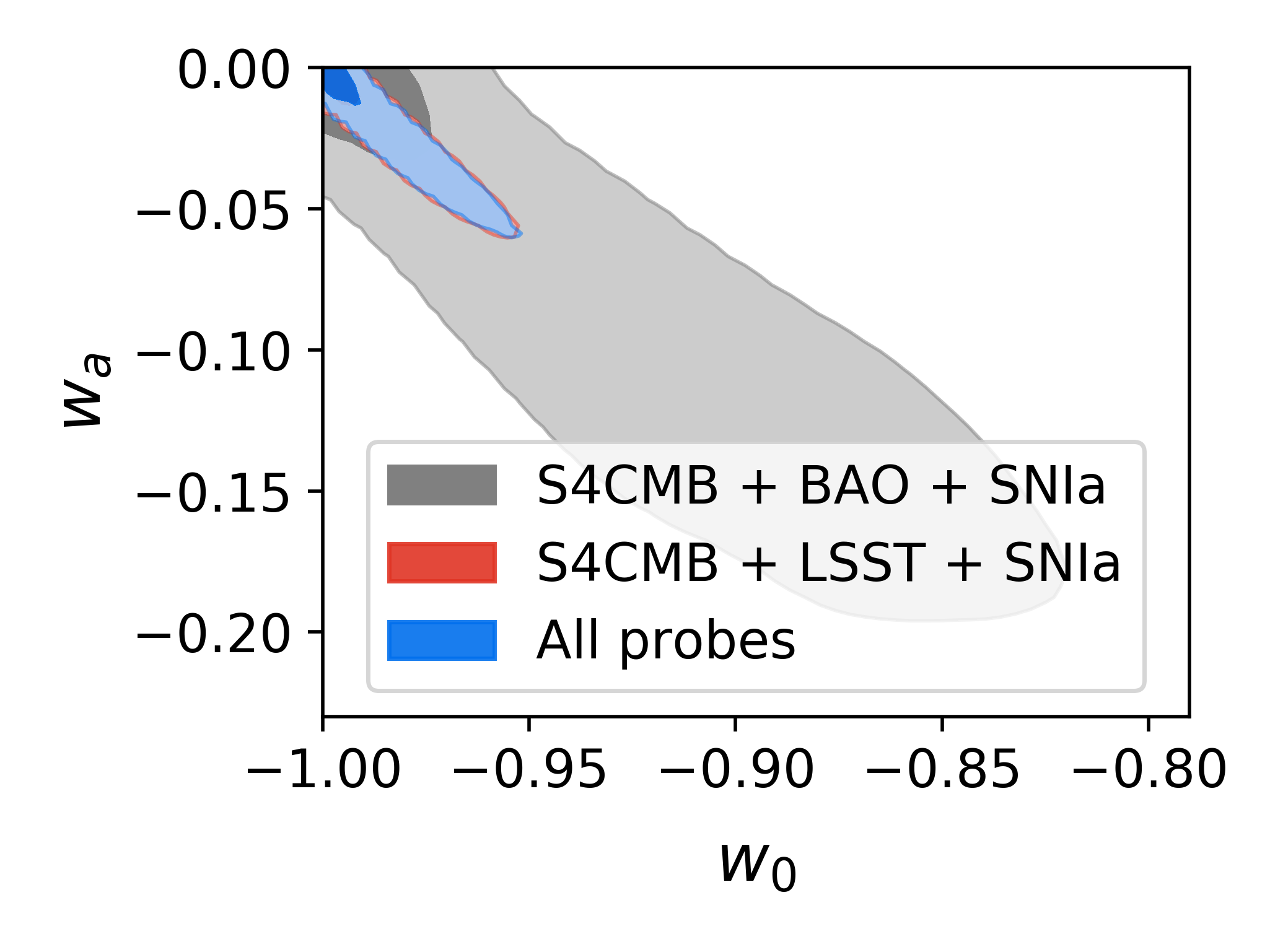}%
  }
  \caption{$2\sigma$-regions of the $w_0-w_a$ parameters parting from a fiducial
    model 1$\sigma$-off the $\Lambda$CDM regime. Note that the maximum constraints are already found for S4CMB
    + LSST + SNIa.}
  \label{fig:w0wa2}
\end{figure}

The fiducial model with maximum is given by $\{\alpha,\, p,\, n\} = \{2,\, 2,\,
3.5\}$. Given that the potential is not symmetric around the maximum, we will
study the forecast potential of the next generation experiments with two fiducial
models with initial value of the field so that it is at both sides of
the maximum. It is located at $\psi_{max} = 1.04$, and we will consider the
cases with $\psi_{ini} = 0.8, 1.4$. The results are shown in
Table~\ref{tab:results}, and the quantitative measurement of the constraining
power of each probe is shown in Table~\ref{tab:FoM}. The found contours are
shown in Fig.~\ref{fig:params2}. As before, galaxy
clustering will be the most constraining probe. In comparison, the case with
$\psi_{ini} < \psi_{max}$ is better constrained, with S4-CMB experiments
having a $\mbox{FoM} \sim 10^{-1}$ and, in combination with BAO and/or SNIa,
$\mbox{FoM} \sim 1$. Using all probes, one can achieve a $\mbox{FoM} \sim
10^2$. However, for the case with $\psi_{ini} > \psi_{max}$, we only reach 
$\mbox{FoM} \sim 1$, when using all probes. The
asymmetry around the maximum is such that at lower values, the potential slope
is much more pronounced (see Fig.~\ref{fig:V-w}), making the model more
sensitive to parameters changes. On the contrary, at values of the field
greater than the maximum, the potential is softer and asymptotically flat,
allowing for greater changes on the parameters that do not impact the final
observables. The greater steepness of the potential is also the reason why the
case with $\psi_{ini} < \psi_{max}$ is more constrained than the case without
maximum (see Fig.~\ref{fig:V-w}), even though the dark energy equation of
state of the fiducial model with maximum is closer to $w=-1$ (see
Fig.~\ref{fig:w0wa2}), as a slightly lower $\psi_{ini}$ would make the field
end up oscillating fast around $0$. It must be noted, however, that it is
still $2\sigma$-off the exact $w=-1$.


As in the previous section, the
main restriction on the dark energy CPL parameters comes from being a thawing
model. In particular, when $\psi_{ini} < \psi_{max}$, the field cannot start
at much lower values than the fiducial $\psi_{ini} = 0.8$, as the field would
roll fast towards $\psi=0$. On the other hand, the constraints allow values of
$\psi_{ini}$ that are closer to the maximum and the plateau, in the case with
$\psi_{ini} > \psi_{max}$. As a consequence, the most likely parameter
combinations that produce a correct late-time acceleration would be those with
$w\sim-1$. Finally, the broader $2\sigma$-contours in the $\psi_{ini} < \psi_{max}$
case, despite of having a $\mbox{FoM} \sim 100$, over the model parameters,
are a consequence of the larger range of accessible values of $V'$ (see the
potential shape in Fig.~\ref{fig:V-w}), which allows a richer variety of field
evolutions. In addition, in all three cases, galaxy clustering is able to
increase the $\mbox{FoM}(w_0,\, w_a)$ by almost an order of magnitude (see
Table~\ref{tab:FoM_w0wa}). The found contours have been plotted in
Fig.~\ref{fig:w0wa2}

The FoM of the model and CPL parameters reflect the fact that the
  phenomenology of this model is mainly determined by its thawing nature and
  the initial position of the field, which determines what part of the
  potential is going to control the field evolution, and not all its
  parameters. In particular, the case with largest FoM on the model parameters
  is that with $\psi_{ini} < \psi_{max}$, while it is the one with lowest
  $\mbox{FoM}(w_0,\, w_a)$.  Similarly, the configuration with $\psi_{ini} >
  \psi_{max}$ has the lowest FoM on the model parameters, but the greatest for
  the CPL parametrization.  Finally, although the case without maximum has a
  $\mbox{FoM}(w_0,\, w_a)$ of same order as the former, its
  $\mbox{FoM}(\psi_{ini},\, \alpha,\, p,\, n)$ is an order of magnitude
  larger. Therefore, this shows the actual degrees of freedom, those that
  affect the phenomenology, are less than the number of free parameters; which
  we already know are degenerated. As a consequence, the
  $\mbox{FoM}(\psi_{ini},\, \alpha,\, p,\, n)$ is not a good quantity to
  inform us about how well constrained is the phenomenology of this model.


\section{Comparison with previous results}

Future observations will be able to greatly constrain the $\alpha$-attractor
model, provided that the true dark energy model were different from a
cosmological constant and $\alpha$ could not be arbitrarily large (i.e. $r
\lesssim 0.01$). In this case, we have shown that a combination of S4CMB +
LSST + SNIa, will greatily improve present results. In fact, they 
increase by almost an order of magnitude the FoM of both the parameter space
and the $w_0-w_a$ parameters, when compared with S4CMB + BAO + SNIa.

\begin{figure}[htb]
  \centering
  \includegraphics[width=0.6\textwidth]{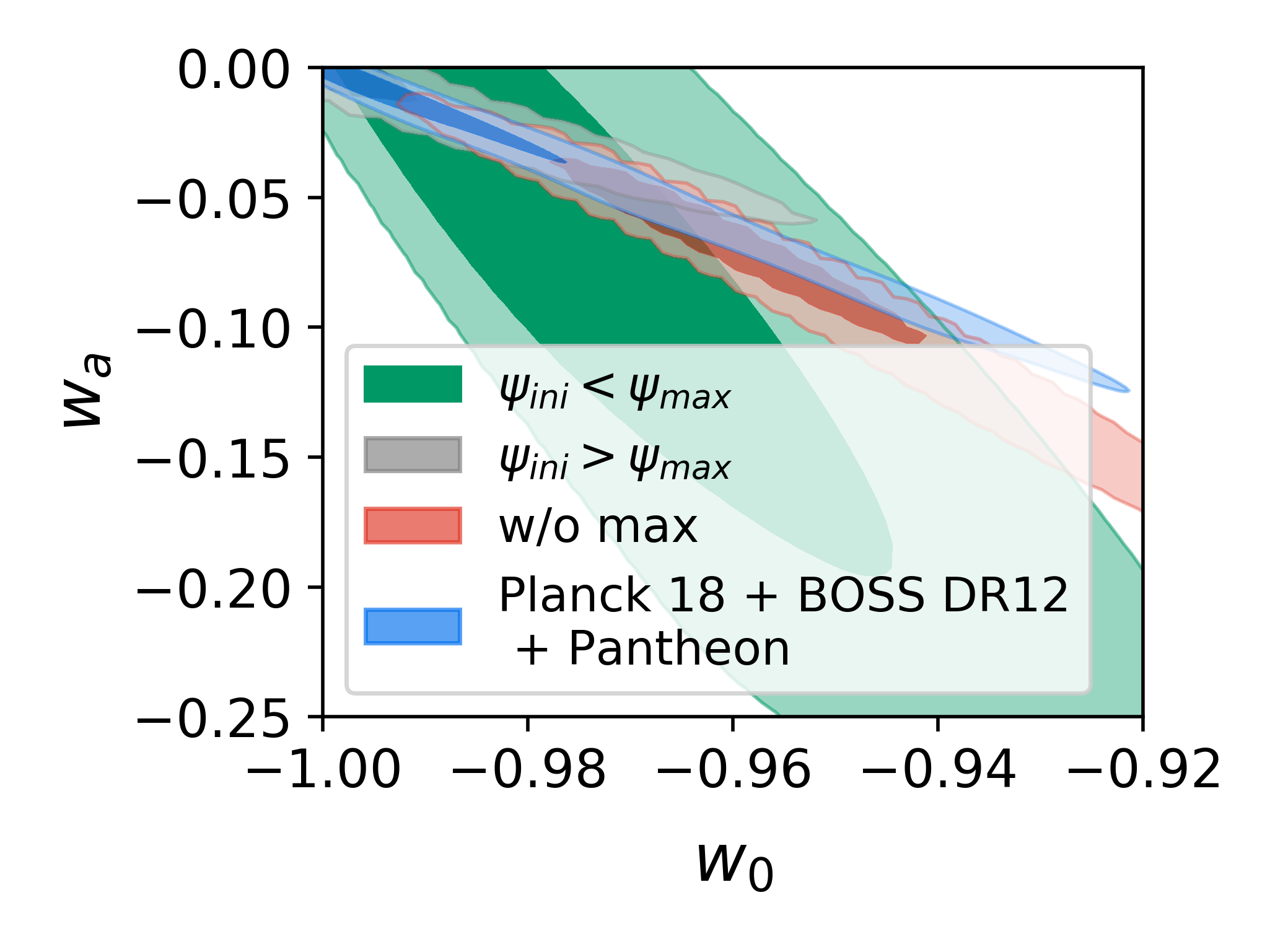}
  \caption{$w_0 - w_a$ $2\sigma$ predicted regions when using all probes,
    compared with the result found in Ref.~\cite{Garcia-Garcia:2018hlc}. The
    case without maximum is $3\sigma$-off a cosmological constant; while the
    case with $\psi_{ini} < \psi_{max}$ is $2\sigma$-off. Forecast constraints
    are broader than those imposed by current observations
    \cite{Garcia-Garcia:2018hlc} for various reasons. First, we are
    performing the forecasts around a specific fiducial cosmology, in contrast to
    the random sampling done in Ref.~\cite{Garcia-Garcia:2018hlc}, where they
    explored all the parameter space without restriction. This shows that
    current observations favored $\Lambda$CDM, which is easily recovered in
    this model (Section \ref{S:model}). In addition, a fiducial model $1\sigma$-off
    a cosmological constant is being used for the forecast analysis. Finally,
    for the particular case with $\psi_{ini} < \psi_{max}$, the $w_0-w_a$
    parameters are not accurate descriptors of the equation of state, as
    equivalent cosmological evolutions can be obtained if $w$ slowly varies
    since early times than in the case it remains close to $w=-1$, but close
    to the present greatly diverges. Nevertheless, the resulting FoM
      for the $\psi_{ini} > \psi_{max}$ case ($8 \times 10^3$) is larger than
      that with current data ($\sim 5\times10^3$) and for the case without
      maximum it is still of the same order of magnitude ($2 \times 10^3$).}
  \label{fig:w0wa_final}
\end{figure}

A special comment is required for the results in the $w_0-w_a$ plane. In
Figure~\ref{fig:w0wa_final} we have plotted the $w_0-w_a$ best
$2\sigma$-contours, together with the results found in
Ref.~\cite{Garcia-Garcia:2018hlc}. The available space for $w_0-w_a$ greatly
depends on the fiducial cosmology used. For instance, if $\psi_{ini} < \psi_{max}$, the
parameters are much less constrained. As we discussed in the previous section, this
is caused by the fact that $V'$ can have a broader range of values that will
modify the acceleration of the field and, in turn, the evolution of the
equation of state. In addition, it also shows that the $w_0 - w_a$ CPL parametrization
is not sufficient to describe the full evolution of the equation of state. In fact, viable and
equivalent cosmologies can be obtained if the equation of state remains $w=-1$
for most of its evolution but grows fast close to the present or in case the
equation of state diverged from $w=-1$ at early times but had a more
monotonically growth along time. The other two cases are more restricted as
the shape of the potential is softer and allows slower evolutions.

The case without maximum is $3\sigma$-off a cosmological constant solution;
while the $\psi_{ini} < \psi_{max}$ is $2\sigma$-off. The case with
$\psi_{ini} > \psi_{max}$ is concordant with $w=-1$ and is caused by the fact
that a mild evolution of the field is allowed given the steepness at that side of the
maximum (see Fig.~\ref{fig:V-w}) and the possibility of having
$\psi_{ini}$ on the plateau by the loose constraints in the parameter space.
It is important to note that the case with $\psi_{ini} > \psi_{max}$ is the
only for which the constraints beat those imposed by current observations,
which have a $\mbox{FoM} \sim 5 \times 10^3$; although, given the mild
constraints we have found in the parameter space, comparing the order of
magnitude is a more conservative approach. This would be the case for the
model without maximum. The reason why our result do not reduce the uncertainty
in the $w_0-w_a$ CPL parameters is caused by the fact that the constraints
that come from current observations (in blue in Fig.~\ref{fig:w0wa_final}) are
showing the preference of current data for a cosmological constant solution as
this can be easily recovered thanks to the degeneracies of this model (see
Section~\ref{S:model}). In comparison with our current approach, the 
constraints from Ref.~\cite{Garcia-Garcia:2018hlc}  were obtained by random sampling in the full parameter space,
with non-informative priors. On the contrary, the constraints found in this
work assume a fiducial model $1\sigma$-off a cosmological constant, and do not
allow the parameters to change the case of study (e.g. $p < n$ when studying
the case without maximum), limiting the possibility of going to the
cosmological constant-like regime.


\section{Conclusions}
\label{S:conclusions}
In this work, we analyze the $\alpha$-attractor dark energy model
\cite{Linder:2015qxa} in the context of near-future cosmological experiments.
This model was already studied with current observations
in Ref.~\cite{Garcia-Garcia:2018hlc} and seen to be unbounded, as a
consequence of the existing parameter degeneracies; in particular,
between $\alpha-\psi_{ini}$ and $p-n$, the potential exponents.

Next-generation experiments will be able to measure the cosmological
observables with percent-level precision. For the specific case with a
maximum ($p < n$) and $\psi_{ini} < \psi_{max}$, we have found an important
improvement on the constraints with respect to current bounds. However,
this improvement does not translate into a significant reduction of
uncertainties in the equation of state parameters under the CPL parametrization. 
This is due to the restrictions of the model in this space of parameters.
On the other hand, the case with $\psi_{ini} > \psi_{max}$ is almost
insensitive to the additional constraining power of next-generation datasets.
Interestingly, in case that true underlying model were that without maximum,
and sufficiently distinct from $\Lambda$CDM, one could detect a $3\sigma$ deviation from a
pure cosmological constant; and a $2\sigma$ deviation if $\psi_{ini} <
\psi_{max}$.

The use of CMB-S4 and other future CMB experiments to place constraints
on the tensor-to-scalar ratio, $r$, and the spectral index, $n_s$, to constrain
$\alpha$ (see Eq.~\ref{eq:alpha}), is unlikely to provide any significant improvement
over the results shown here, since those constraints will still allow for too much
freedom, leaving the results shown in Ref.~\cite{Garcia-Garcia:2018hlc} almost untouched.

Finally, the use of tomographic galaxy clustering would be particularly
important in order to achieve this. From the analysis of individual probes
(see the Figures of Merit in Table~\ref{tab:FoM}), we have shown that galaxy
clustering will be the probe with the most constraining capability, since the
galaxy power spectrum is the most sensitive observable
(Fig.~\ref{fig:derivatives}) to changes in the $\alpha$-attractor parameters.
We find this statement to be true across the different fiducial models
studied. In particular, the combination S4CMB + LSST + SNIa will improve the
FoM of both the parameter space and $w_0 - w_a$ by almost an order of
magnitude with respect to the case with S4CMB + BAO + SNIa.

Next-generation experiments will lead us to an unprecedented level of
precision in cosmology, allowing us to test our knowledge about the Universe,
its origin and dynamical evolution. In this work we have shown how these
observations, in particular a combination of CMB, galaxy and SNIa measurements
will be able to set constraints on the dark energy $\alpha$-attractor model
and, as a consequence, we would expect that, in general, future surveys will
be able to probe whether the late accelerated expansion of the Universe is
connected with the one the Universe started with -- inflation.

\begin{acknowledgments}
  We would like to thank Emilio Bellini, Eva-Maria Mueller and Pedro Ferreira for useful
  comments and discussion. CGG and PRL are supported by AYA2015-67854-P from the Ministry
  of Industry, Science and Innovation of Spain and the FEDER funds. CGG is supported by the
  Spanish grant BES-2016-077038, partially funded by the ESF, and was
  partially supported by a Balzan Fellowship while in Oxford. He would like to
  thank New College and the Department of Physics at Oxford, as well as BCCP,
  LBLN and UC Berkeley for their hospitality. DA acknowledges support from
  STFC through an Ernest Rutherford Fellowship, grant reference ST/P004474/1.
  MZ is supported by the Marie Sklodowska-Curie Global Fellowship Project
  NLO-CO.
\end{acknowledgments}



\bibliography{alpha_attractors-forecast}
\bibliographystyle{JHEP}

\end{document}